\documentclass{article}
\usepackage{amsmath,graphicx,url,times,authblk}

\RequirePackage[]{algorithm2e}
\RequirePackage{multirow}
\RequirePackage[justification=centering]{caption}
\RequirePackage{subcaption}

\title{ABROA : Audio-Based Room-Occupancy Analysis using Gaussian Mixtures and Hidden Markov Models}

\author[1]{Rafael Valle\thanks{This research was supported in part by the TerraSwarm
Research Center, one of six centers supported by the STAR net phase of the Focus
Center Research Program (FCRP) a Semiconductor Research Corporation program
sponsored by MARCO and DARPA.}}

\affil[1]{UC Berkeley,
         Center for New Music and Audio Technologies (CNMAT),\\
         Berkeley, California, 94709, USA\\
         rafaelvalle@berkeley.edu
        }

\begin{document}
\maketitle

\begin{sloppy}

\begin{abstract}
This paper outlines preliminary steps towards the development of an audio-based room-occupancy analysis model. Our approach borrows from speech recognition tradition and is based on Gaussian Mixtures and Hidden Markov Models. We analyse possible challenges encountered in the development of such a model, and offer several solutions including feature design and prediction strategies. We provide results obtained from experiments with audio data from a retail store in Palo Alto, California. Model assessment is done via \textit{leave-two-out} Bootstrap and model convergence achieves good accuracy, thus representing a contribution to multimodal people counting algorithms.

\end{abstract}

%\begin{keywords}
%    \input{keywords}
%\end{keywords}

\section{INTRODUCTION}\label{sec:intro}
    Information about the occupancy of a certain location is relevant for several
applications, specially in surveillance tasks and staff management. For
example, occupancy can be used to detect intruders in a house, or to generate
optimal employee schedules according to shopper traffic. The existing systems
for occupancy detection rely on multimodal systems, including video, Wifi,
Bluetooth and, to a much lesser extent, audio. Conversely, the use of audio in
occupancy estimation empowers the development of a model, not dependent on
speaker separation, that is robust to issues that are common in computer vision and systems that rely on tracking electronic devices\footnote{An increasingly hard task with the randomization of MAC addresses and privacy laws}, such as occlusion and people clusters\cite{moeslund2001survey}.

\subsection{Related work}
The occupancy estimation literature can be subdivided in invasive and non-invasive strategies. Invasive strategies~\cite{srinivasan2012presence,xu2013crowd++,shih2015occupancy} use various devices, e.g. smartphones and ultrasonic transmitters, to project sound, e.g. sinusoids and chirps, onto the environment and use the environment's response to the projected sound to estimate activity in the space; Non-invasive strategies~\cite{khan2015infrastructure,kannan2012low,stillman2001towards} rely on detecting speech sounds in the environment to estimate occupancy.
In addition to potentially disturbing humans and animals, invasive strategies require the expensive task of deploying devices, e.g. 891 mobile devices to cover 600 square meters and less than 20 people~\cite{kannan2012low}, in the location and badly suffer from the addition of non-human objects and subjects to the space being analyzed.
Non-invasive strategies that rely on speech only to estimate occupancy will
disregard people who are in a space but not talking, and badly suffer from
situations in which speech diarization is not possible. In addition, most of these systems are only
able to handle small groups of people.

The limitations presented above and the lack of a standard technique or key
paper on the topic of audio-based room-occupancy analysis confirm the need for
the development of such a technique. There is no annotated dataset for
audio-based room-occupancy analysis and, therefore, the acquisition of data
remains a blatant challenge. Although~\cite{uziel2013networked} describes the layout of a rather promising
prototype to estimate the occupancy of rooms and buildings based on audio, they
provide no information on experiments nor results describing the efficiency of
their system. In this paper we describe a system that is not invasive, suited for large groups of people, based on real data and computationally 
inexpensive.

\section{METHODOLOGY}\label{sec:methodology}
    \subsection{Dataset and ground truth}
The dataset used in this research is comprised of proprietary audio recordings made with a smartphone placed in a retail store located in the United States. The smartphone's microphone was aimed towards the inside of the store and placed at the store's main and single entrance, at approximately 5 meters from the floor. The recordings took place during open hours (10h, 22h).
The ground truth data is divided into 15-minute slices and it provides the cumulative occupancy at the end of each 15-minute time window. The ground truth was obtained from video data submitted to Amazon's Mechanical Turk. Figure \ref{fig:occupancy} illustrates the occupancy for that specific week. We invite the reader to consider the distribution of occupancy.

\begin{figure}[h!]
    \centering
    \caption{Occupancy}
    \includegraphics[width=0.8\linewidth]{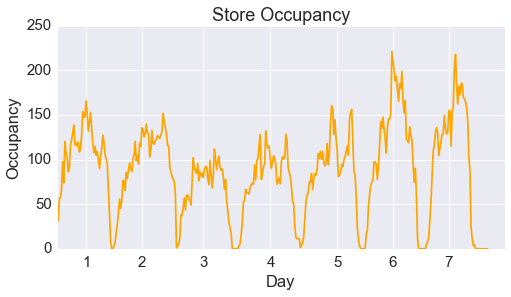}
  \caption{Store occupancy for the first week of April.}
  \label{fig:occupancy}
\end{figure}

\subsection{Room-occupancy Analysis}
Several challenges are present in the development of an audio-based
room-occupancy analysis system. In our context, the ground truth only provides
information about the aggregated occupancy at the end of each 15-minute
interval. This provides a challenge to feature selection, that is, selecting the
audio slice that best represents the ground truth. In addition, since there's
one dependent variable for each 15-minute interval, regression models would
require the design of summary statistics of the audio data to be used as the
independent variable, thus extremely reducing the amount of information
retrieved from each training sample. During evaluation we considered generalized
linear models (GLM). Surprisingle, the mean error of the best linear model, Poisson Regression, was
33\% worse than the GMM-HMM model described in this paper.

\subsubsection{Audio Features}
In our experiments we performed cross-validation on the training set using lasso regression models with the following features and linear regression models using all posible combination of the following features:
\begin{description}
\item [amplitude] median, mean, standard deviation
\item [spectral] centroid, spread, skewness, kurtosis, slope
\item [mfcc] raw, 1st delta, 2nd delta
\end{description}

We observed the p-values, 5\% significance, of the regression models and concluded that the MFCC features contributed the most to prediction accuracy on the training set. Given this conclusion, our room-occupancy analysis algorithm uses the well-known Mel-Frequency Cepstral Coefficients\cite{gold2011speech} (MFCC). A total of 20 MFCCs (computed with a FFT Size of 4096 samples, Hop Size of 1024 samples and audio
sample rate is 11050 hz) along with their first (20 features) and second (20
features) deltas\cite{furui1986speaker} assembling a feature vector with 60 dimensions total. Given that sounds produced by humans are rarely stationary, the delta features provide valuable information.

\subsubsection{Window selection}
Our labeled data provides the cumulative occupancy at the end of a 15-minute
time window. This represents two challenges: first, it is necessary to find out
what temporal slice of the audio data should be used; second, the length of
this slice must be chosen such that it maximizes the model's performance. 
Window size is chosen under the one-standard-error rule using 50 iterations of
\textit{leave-two-out} Bootstrap with window sizes in the interval [30, 260]
seconds, with a 10 seconds step size and starting at the end and increasing
towards the beginning of the audio file.

\subsubsection{GMM-HMM model}
We propose a solution that references the speech recognition literature\cite{rabiner1989tutorial} and combines Gaussian Mixtures with Hidden Markov
Models. The GMM is appropriate, for it provides a better categorization of the
distribution of the audio features and a reliable estimate of the likelihood
function, $p(X | \lambda)$, where $X = {x_1, x_2, \ldots, x_T}$ is a sequence of
feature vectors (MFCCs and deltas in our case), $x_t$ is a feature vector
indexed at discrete time $t \in [1,2,...,T]$, and $\lambda$ represents some
model. As described in\cite{reynolds2000speaker}, for a D-dimensional feature vector $x$
(60-dimensional in our case), the mixture density used for the likelihood of
data $x$ given model $\lambda$ is defined as:
\begin{equation}
  p(x | \lambda) = \sum_{i=1}^{M}w_ip_i(x)
\end{equation}
This density is a weighted linear combination of $M$ unimodal Gaussian densities, $p_i(x)$, with parameters $\mu_i$ ($D \times 1$ mean vector) and $\Sigma_i$ ($D\times D$ covariance matrix):

\begin{equation}
  p_i(x) = \frac{1}{(2\pi)^{D/2}|\Sigma_i|^{1/2}}\exp\{-\frac{1}{2}(x - \mu_i)'\Sigma_i^{-1}(x-\mu_i)\}
\end{equation}

Under the assumptions in\cite{reynolds2000speaker}, our model only uses the
diagonal covariance matrix and the maximum likelihood model parameters are
estimated using the well-known iterative expectation-maximization (EM)
algorithm\cite{rabiner1989tutorial}. Traditionally, the feature vectors of $X$ are assumed independent and the log-likelihood for some sequence of feature vectors X is computed as:
\begin{equation}
    LL(X|\lambda) = \sum_{t=1}^Tlog(p(x_t|\lambda))
\end{equation}
We bin our occupancy data by taking the integer square root of occupancy values, thus circumscribing the problem of creating one GMM per occupancy value. The lowest occupancy is 0 and the maximum is 221, thus producing 15 occupancy bins. For each occupancy bin and its respective audio data, one bin-dependent GMM is trained with the MFCC features described above and using the Bayesian Information Criterion (BIC) for model selection over the number of components in the set $C =
\left\{2,3,4,5,6,7,8,16,32\right\}$ on a test set.

In addition to the GMM, a HMM\cite{rabiner1986introduction} can be used to compute $P(O|\lambda)$, that is the probability of the observation sequence $O = o_1, o_2, \dots, o_T$, i.e. occupancy sequence, given model $\lambda$. As described in\cite{blunsom2004hidden}, the probability of observations $O$ for a fixed state sequence $Q = q_1, q_2, \dots, q_T$, $P(O|Q,\lambda)$ is:
\begin{equation}
  \prod_{t=1}^{T} P(o_t|q_t, \lambda) = b_{q_1}(o_1) b_{q_2}(o_2) ... b_{q_T}(o_T)
\end{equation}
where $b$ is an array storing the emission probabilities at time $t$ given model (bin-dependent GMM in our case). The probability of the state sequence $Q$ is given by:
\begin{equation}
  P(Q|\lambda) = \pi_{q_1}a_{q_1q_2}a_{q_2q_3}...a_{q_{T-1}q_T}
\end{equation}
where $\pi$ is an array storing the initial probabilities and $a$ is an array storing the transition probability from state $q_{t}$ to state $q_{t+1}$. We can calculate the probability of the observations given the model as:
\begin{equation}
  P(O|\lambda) = \sum_Q P(O|Q,\lambda)P(Q|\lambda)
\end{equation}
Decoding of the hidden state is computed using the Viterbi\cite{forney1973viterbi} algorithm to find the best path (single best state sequence) for an observation sequence. We define the probability of the best state path for the partial observation sequence as:
\begin{equation}
  \delta_t(i) = \max_{q_1, q_2, \cdots, q_{t-1}} P(q_1q_2, \cdots, q_t = s_1, o_1, o_2, \cdots, o_t | \lambda)
\end{equation}
Figure \ref{fig:hmm} illustrates our system and its use of the Hidden Markov Model and the Viterbi algorithm. Each circle represents the log-likelihood score of each feature vector given each binned GMM. The lines represent the transition probabilities and the highlighted states and bold lines show the state path that maximizes the log-likelihood.
\begin{figure}[h!]
  \centering
  \includegraphics[scale=0.26]{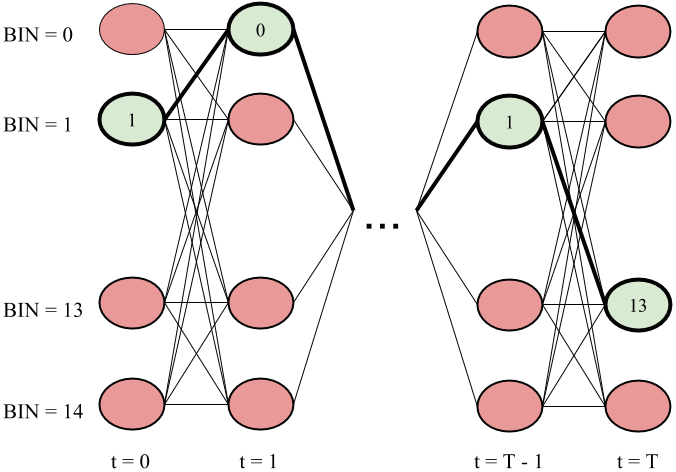}
  \caption{HMM and Viterbi illustration}
  \label{fig:hmm}
\end{figure}

\section{EXPERIMENTS AND RESULTS}\label{sec:experiments}
    Several techniques were used for prediction, starting with using the log-likelihoods of each feature vector given each GMM and ending with a HMM.
\subsection{Prediction with GMMs}
Following the bin-dependent GMM training, two prediction strategies are chosen, including aggregating the posterior probabilities of each GMM and majority voting.

\subsubsection{Posterior Probabilities Aggregation (PPA)}
This procedure consists of aggregating the posterior probabilities of each state by computing the sum of bin occupancy predictions weighted by their posterior probabilities. Let $x$ be an audio feature vector and $\Lambda$ be a set of 15 bin-dependent GMMs. From Bayes theorem:
\begin{equation}
P(x|\Lambda_i) = \frac{P(\Lambda_i | x) P(x)}{P(\Lambda_i)}
\end{equation}
Assuming that $P(\Lambda)$ is uniform and knowing that $P(x)$ is similar for all models, we conclude that $P(\Lambda_i|x) \propto P(x|\Lambda_i)$. Finally, we define the estimated occupancy bin ($\hat{B}$)\footnote{We use log-sum-exp to prevent computational underflow} for feature vector $x$ at time $t$ and models $\Lambda$ as:
\begin{equation}
  \hat{B}(x_t) = \sum^{|\Lambda|-1}_{i=0} ie^{LL(x_t | \Lambda_i)}
\end{equation}

\subsubsection{Majority Voting (MJ)}
This procedure consist of calculating the log-likelihood, $LL(X, \lambda)$ of the audio features given each bin-dependent GMM and finally selecting the bin that more often has the highest log-likelihood score.

Using the GMM only approach and these techniques, informally speaking the prediction results circulate around the correct prediction value. However, the results for both techniques show jumps in occupancy prediction that are rather unlikely because the model ignores transition probabilities between states. Therefore, we decided to address this problem by using a HMM and the
Viterbi algorithm.
\if 0
\begin{figure}[h!]
  \begin{subfigure}[t]{0.35\textwidth}
    \centering
    \caption{PPA}
    \includegraphics[width=\linewidth]{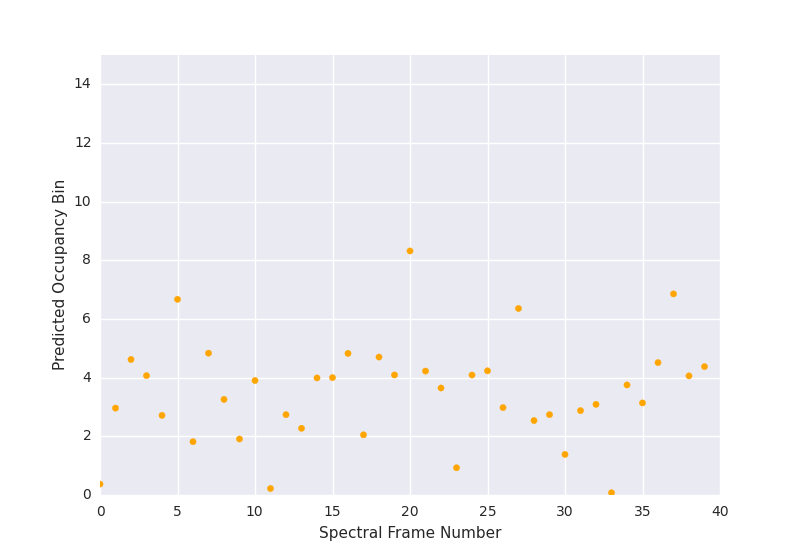}
  \end{subfigure}
  \begin{subfigure}[t]{0.35\textwidth}
    \centering
    \caption{MJ}
    \includegraphics[width=\linewidth]{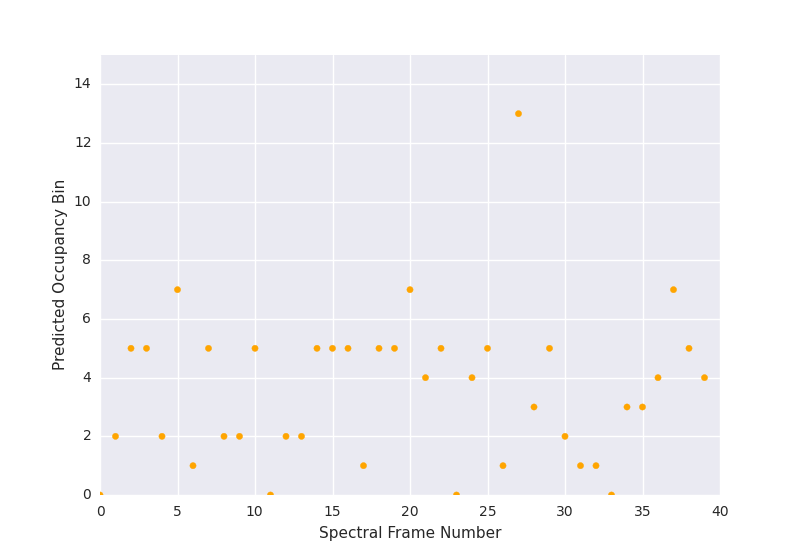}
  \end{subfigure}
  \caption{Predictions with GMM. On average, the PPA technique underpredicts occupancy and the MV technique spreads its prediction between bins 2 and 5, with 5 being the winning vote.}
  \label{fig:gmm_agg_pred}
\end{figure}
\fi
\subsection{Prediction with HMM and Viterbi}
This strategy divides prediction into four stages, as described in Algorithm \ref{algo:prediction}. The first computes the log-likelihood of the audio features given each bin-dependent GMM; the second applies the Viterbi algorithm to the computed log-likelihoods to obtain the best path for a specific time window. The third predicts the occupancy bin using posterior probabilities aggregation or majority voting; finally, the inverse of the square root of the predicted bin is calculated,
thus translating the prediction back into the linear domain. For the HMM, the
initial probabilities $\pi$ are uniform (1/15), the emission probabilities $b$
are computed using the GMMs and the transition probabilities $a$ are derived
using heuristics.

\begin{algorithm}
  \caption{Occupancy Prediction}\label{OP-Algorithm}
  \label{algo:prediction}
  \KwData{audio features}
  \KwResult{predictions}
   GMMs $\gets$ Train Bin-dependent GMMs on features\\
  \For{feature in audio features} {
    LLs $\gets$ computeLL(feature, GMMs)\;
    bestpath $\gets$ viterbi(LLs)\;
    prediction $\gets$ PPAorMV(bestpath)\;
    predictions $\gets$ [predictions, prediction$^2$]\;
  }
\end{algorithm}
\if 0
Figure \ref{fig:conf_mat_training} illustrates the results obtained using the algorithm \ref{algo:prediction} above, with training and prediction using all data.
\begin{figure}[h!]
  \centering
  \begin{subfigure}{\linewidth}
    \centering
    \caption{Posterior Probabilities Aggregation}
    \includegraphics[scale=0.3]{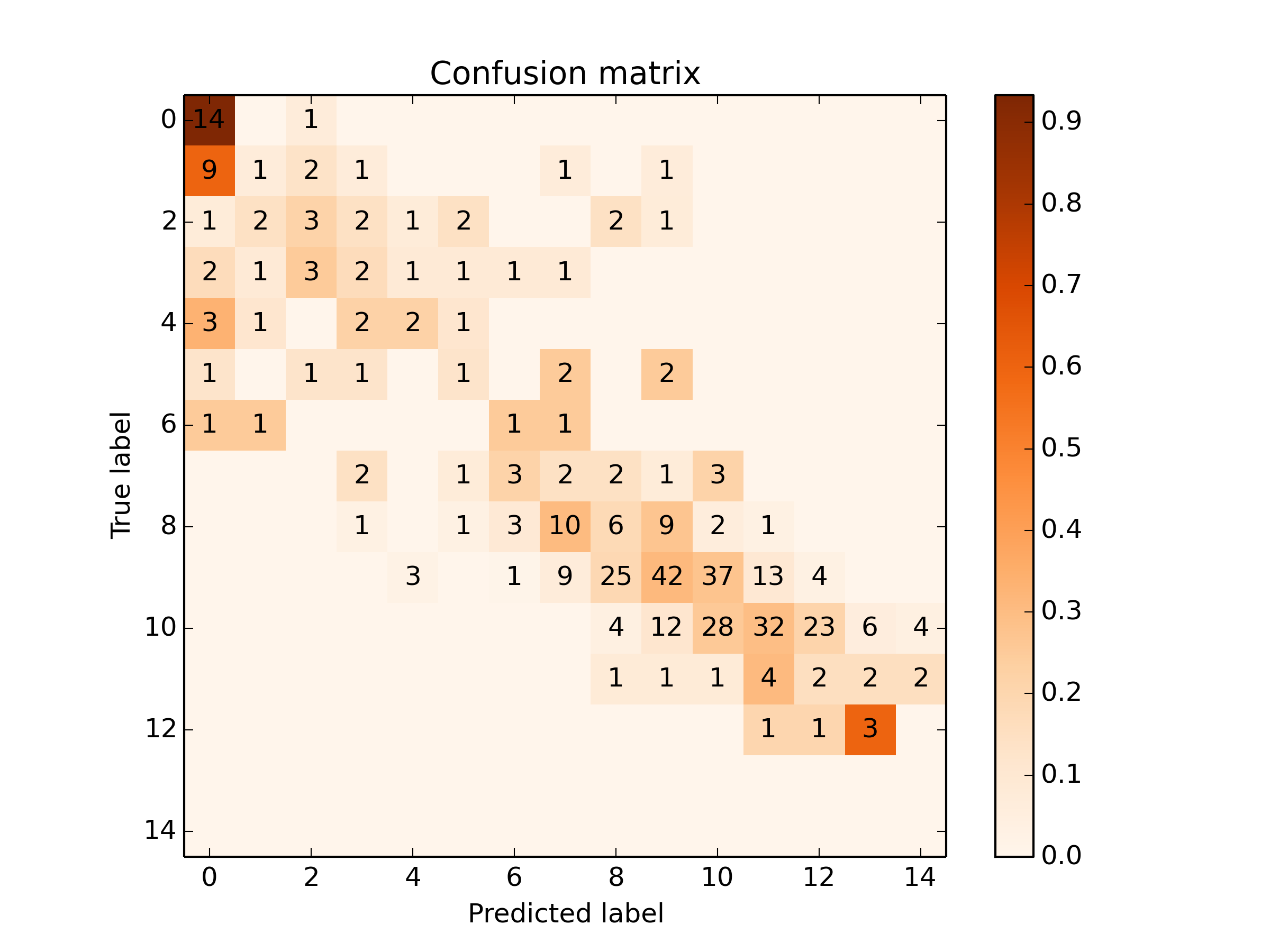}
  \end{subfigure}
  \begin{subfigure}{\linewidth}
    \centering
    \caption{Majority Voting}
    \includegraphics[scale=0.3]{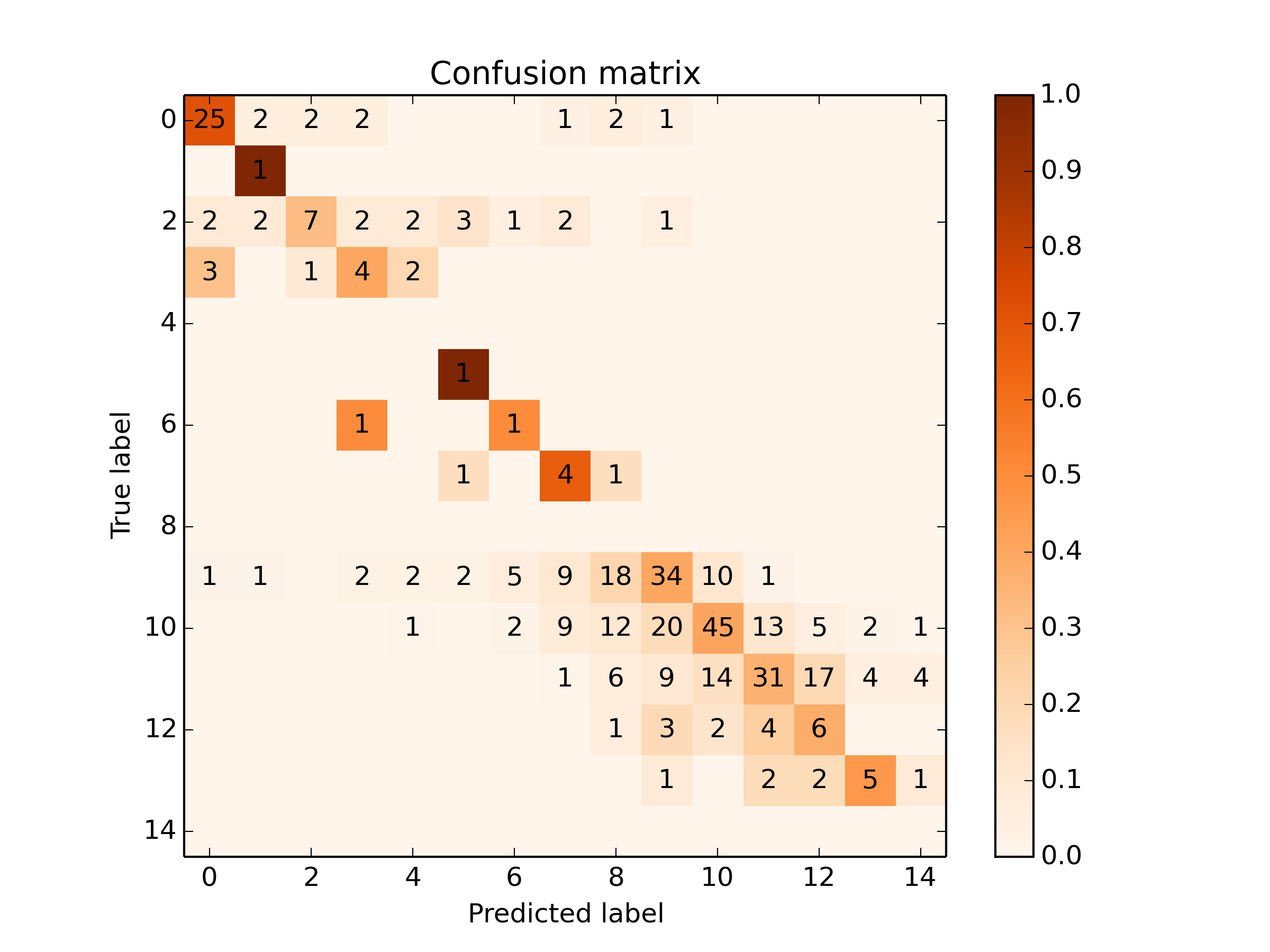}
  \end{subfigure}
  \caption{Confusion Matrices for the PPA and MV prediction techniques using HMM with the Viterbi algorithm}
  \label{fig:conf_mat_training}
\end{figure}

Clearly the posterior probability aggregation technique overpredicts occupancy, as can be seen by the vertically shifted confusion matrix on figure \ref{fig:conf_mat_training}. On the other hand, the majority voting technique produces a clear diagonal line that indicates better prediction results, despite the
\fi
\subsection{Model Assessment and Selection}
We used \textit{leave-two-out} Bootstrap\cite{hastie2009elements} for model assessment and selection, as it is a suitable  technique for our small dataset of 386 samples. The Bootstrap technique was used to estimate the best window size, within the range [30, 260] seconds, and the most accurate prediction strategy based on the HMM-Viterbi model. For each window size and prediction strategy, a total of 50 bootstrap iterations were performed and the prediction errors were computed.

Figure \ref{fig:bootstrap_window} shows bootstrapped root mean squared errors
(RMSE) for all window sizes and using the MJ and PPA techniques. Accuracy
increases, specially in PPA, almost linearly and proportionally to the window size.
\begin{figure}[h!]
  \centering
  \includegraphics[width=\linewidth]{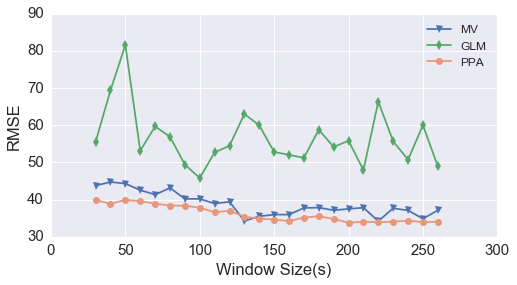}
  \caption{RMS errors for different windows and prediction techniques using
    HMM-Viterbi and Poisson Regression(GLM). The PPA technique converges around 210 seconds 
    and considerably outperforms the MV and GLM techniques.}
  \label{fig:bootstrap_window}
\end{figure}

Model and window selection is performed using the one-standard-error rule and analysis of the violin plot provided in Figure \ref{fig:bootstrap_violin}. The best predictor uses the PPA technique with a window size of 210 seconds. 
\begin{figure}[h!]
  \centering
  \includegraphics[width=0.9\linewidth]{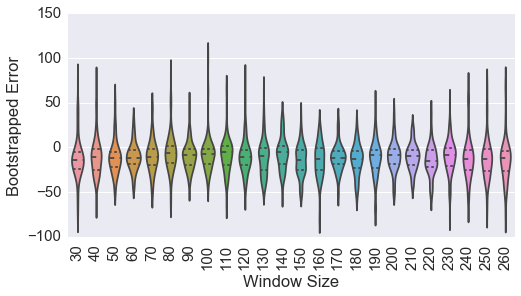}
  \caption{Bootstrapped error using GMM-HMM and PPA. Overall the bootstrapped error is below zero, thus suggesting that the model on average underpredicts.}
  \label{fig:bootstrap_violin}
\end{figure}

Figure \ref{fig:bootstrap_prediction} shows room-occupancy predictions with a window of 210 seconds and both prediction techniques. Although the predictions made by both strategies closely follow the ground truth's profile, there are individual large errors such as around sample indices 10 and 80.

\section{CONCLUSION AND FUTURE WORK}\label{sec:conclusions}
    We have described an algorithm for audio-based room-occupancy analysis that relies on Gaussian Mixtures and Hidden Markov Models. Our algorithm has advantages over other algorithms for audio-based occupancy analysis:
\begin{description}
  \item[It is not invasive] our algorithm does not require projecting audio into an environment, thus not causing disturbance to humans or animals present.
  \item[It handles large groups] our algorithm is capable of handling occupancy prediction up to 200 people and this capacity can be extended given the appropriate training data.
  \item[It is inexpensive] prediction is computationally cheap and can be easily done on a smart phone, thus preventing privacy issues that might arise if audio data is sent over the network for prediction.
\end{description}
We analyzed different types of prediction techniques and concluded that the GMM-HMM posterior probabilities aggregation is the preferred approach, yielding better results than all other strategies explored. The algorithm performed considerably well in retail store environments with occupancy up to 200 people.

\begin{figure}[h!]
  \centering
  \includegraphics[width=\linewidth]{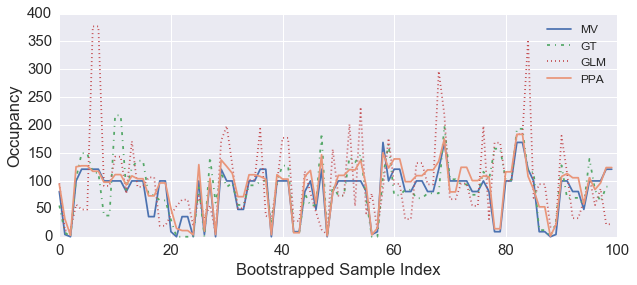}
  \caption{Ground Truth (GT) and predictions using the above described
    techniques (MV and PPA). The model performs well, in spite of two crass errors
    between indices 0 and 20. Unexpectedly, the generalized linear model 
    (Poisson Regression) performs considerably worse then the models proposed in
    this paper.}
  \label{fig:bootstrap_prediction}
\end{figure}

The results from the current work validate the model and justify collecting
more data to build a more balanced dataset with the foresight of increasing
accuracy. In addition, we plan to use occupancy data to create the transition probabilities, instead of relying on ad-hoc rules. Last, we plan to add a Voice Activity Detection pre-processing step to the pipeline, denoise the audio data and perform comparative analysis with the results obtained in this paper.

\if 0 
\section{ACKNOWLEDGMENTS}\label{sec:acknowledgements}
    \input{acknowledgments}
\fi

\bibliographystyle{IEEEtran} % style aes.bst
\bibliography{abroa_arxiv}
\end{sloppy}
\end{document}